# An Experimental Study on Turbulent-Stripe Structure in Transitional Channel Flow

Takahiro TSUKAHARA*, Yasuo KAWAGUCHI, Hiroshi KAWAMURA

Department of Mechanical Engineering, Tokyo University of Science.
*E-mail: *tsuka@rs.tus.ac.jp*

**Abstract** ― Turbulent stripe, which would occur in turbulent channel flows at transitional Reynolds numbers, was studied experimentally by flow visualization using reflective flake particles. In a range of bulk mean Reynolds number $Re_m$ = 1700–2000, the turbulent stripe was observed to be inclined at angles of 20°–30° against the streamwise direction, and its streamwise wave length was about 60 times of the channel half width ($\delta$). The longitudinal streaks with the spanwise spacing of $1.6\delta$ were found in the quasi-laminar regions. The critical Reynolds number was $Re_m$ = 1300. Time traces of the streamwise velocity were measured by laser Doppler velocimetry, revealing similarity to the equilibrium turbulent puff in the transitional pipe flow.

## 1. Introduction

The transitional flow has been studied by many researchers for a long time and remains to be an unresolved problem of nonlinear physics, even though much efforts have been devoted to such an important issue with its many applications in different areas of engineering, bio-fluid dynamics and geophysics. One of the outstanding questions pertains to the structures of the transition process, namely the turbulent puffs and slugs. They were firstly observed and classified in a transitional pipe flow by Wygnanski *et al.* [1,2]. While slugs were associated with transition from laminar to turbulent flow, puffs represented an incomplete relaminarization process. Puffs could only be seen at $Re$ = 2000 ~ 2700, while slugs occur at $Re$ higher than 3200. These phenomena involve a mixed state of turbulent patches and ordered laminar domains, which coexist for the same values of the control parameter. This kind of scenario in transition has been observed in different systems with large-aspect ratio such as variety of Couette flows, as described later.

Recently, the equilibrium turbulent structure similar to the puff was found in a turbulent Poiseuille flow at very low Reynolds numbers by Tsukahara *et al.* [3] through a direct numerical simulation (DNS) with huge computational boxes (up to $327.7\delta \times 2\delta \times 128\delta$). Consisting of a turbulent and a laminar region, each with a stripe pattern, turbulent stripe was inclined about $\theta$ = 20°~25° against the streamwise direction. The most energetic wavelengths were about $65\delta$ in streamwise direction and about $26\delta$ in spanwise direction. When the turbulent stripe was occurred, the values of transitional and critical Reynolds numbers below which the turbulent flow became intermittent and laminar were found to be decreased. Therefore it is implied that the turbulent stripe should be effective in sustaining turbulence and in heat-transfer enhancement. The propagation velocity of the turbulent stripe was almost same with the bulk mean velocity. However, although studied numerically further in these years [4-6], the flow field which exhibits the turbulent stripe has not been investigated by experiments. In numerical simulations, the periodic boundary condition, which is usually applied in horizontal directions to assume an infinite channel, would inevitably affect the stripe pattern. Moreover the onset of the turbulent stripe and the Reynolds number below which the pattern appeared spontaneously from homogeneous turbulence were not yet verified by experiments. The critical Reynolds number below which any turbulent disturbance decays is neither proven to be unique. This is one of the important issues to be paid attention.

On the other hand, similar spatiotemporal intermittent states such as the spiral turbulence in Taylor-Couette flow, were studied by a number of researchers. Andereck *et al.* [7] provided the first characterization of the states and the transitions between them in Taylor-Couette flow with reflective flakes. They observed some of the flow states between counter-rotating cylinders including laminar spirals, spiral turbulence, and a flow with intermittent turbulent spots. Hegseth *et al.* [8] observed and measured for the first time a non-uniform pitch in long geometries and its dependence on boundary conditions at the cylinder end and they explained these results within the framework of phase dynamics. The turbulent stripes in a Taylor-Couette and a plane Couette flow were studied experimentally by Prigent *et al.* [9], in which both the transition from laminar to turbulent flow and the reverse transition from turbulent to laminar flow were considered. They reported that the transition to turbulence was discontinuous and characterized by laminar-turbulent coexistence, whereas the reverse transition was continuous and led to a periodical stripes pattern. The Turbulent stripe in plane Couette flow was studied numerically by Barkley *et al.* [10]. They observed the flow in the quasi-laminar region was not the linear Couette profile, but results from a non-trivial balance between advection and diffusion. In addition, they made a comparison of the turbulent stripe in plane Couette flow with the turbulent-laminar banded patterns in other flow systems: plane Poiseuille, Taylor-Couette and rotor-stator. Carlson *et al.* [11] revealed some striking features of turbulent spots in Poiseuille flow by flow visualization of artificially triggered transition with reflective flakes. They had shown that there were strong oblique waves at the front of the arrowhead-shaped spot as well as trailing from the rear tips. Moreover, they showed that strong oblique wave propagation and breakdown play a crucial role in transition to turbulence in Poiseuille flow. In their observation, visualization of the flow state was accomplished using a mixture of reflective flakes. It is known that reflective flakes tend to align with the shear layers when suspended in liquids [7,11]. Because of this, vortex can be seen by flow visualization with the reflective flakes.

In the present study, the structure of turbulent stripe in a transitional water channel flow was investigated by flow visualization with reflective flakes for $Re_m$ = 1300 ~ 2500, and the streamwise velocity was measured by LDV for $Re_m$ = 1100 ~ 2400. The Reynolds numbers of $Re_m$ based on the bulk mean velocity ($u_m$), the channel width $2\delta$ ($\delta$ = 5 mm) and the kinematic viscosity ($\nu$). Moreover, the critical Reynolds number and the shape of the turbulent stripe were examined.

## 2. Experimental Set-up

Figure 1 shows the flow and visualization systems. The experiments were performed on a closed-circuit water loop composed of a straight channel, tanks, a flow meter and a pump. The two-dimensional channel is made of transparent acrylic resin with 4 m length, 0.01 m height and 0.4 m width. A turbulence grid was installed at the entrance of the channel test section. The present Reynolds number was lowered gradually from turbulent flow to laminar flow. To visualize the flow, light reflecting flakes of mica platelets (product of Merck KGaA; product name is Iriodin®123.) were added in the water. A size range of platelets was about 5~25 μm.

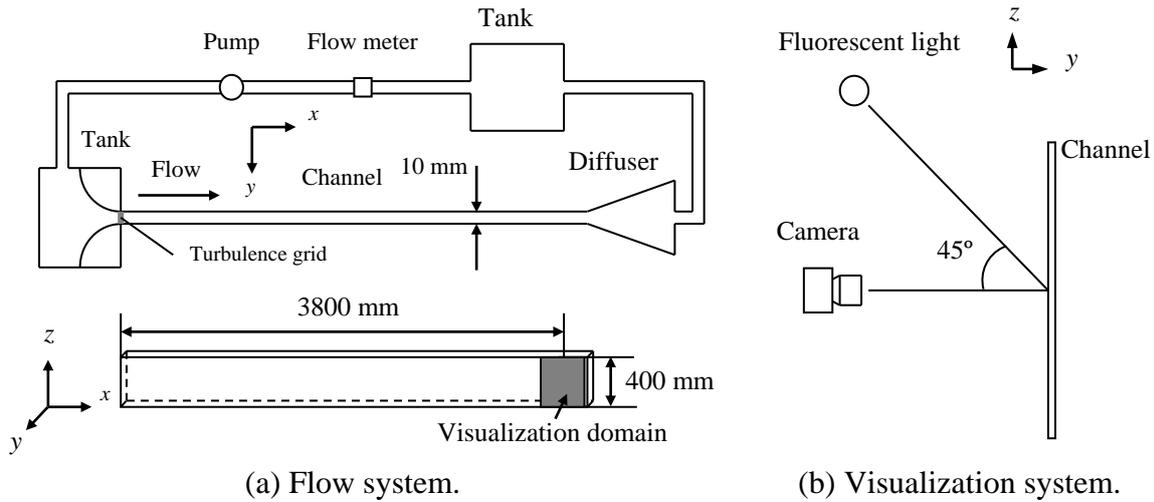

(a) Flow system.   (b) Visualization system.

Figure 1: A schematic diagram of the experimental apparatus and visualization system.

The test section was illuminated by a fluorescent light, and a single-lens reflex camera (Nikon D60) and a video camera (Sony DV Handycam) with CCD were used. The camera has a resolution of 3872 × 2592 pixels. The shutter speed was 1/125 s. We used the wide-angle lens (SIGMA 28mm F1.8 EX DG Aspherical Macro). The still camera was used to record the flow pattern precisely. The video camera has a resolution of 720 × 480 pixels on the frame rate of 29.97 fps. The video camera was used to record and to analyze the evolution of the flow path to identify the space-time structure of the stripe patterns. The fluorescent light was located at the angle of 45º against the optical axis of the camera. Flow visualization was performed for $1300 \leq Re_m \leq 2500$. Using the still camera, the flow visualization domain of $70\delta \times 70\delta$ in an ($x$, $z$) plane was $760\delta$ downstream from the turbulence grid. Using the video camera, the flow visualization domain of $126\delta \times 84\delta$ in an ($x$, $z$) plane was $760\delta$ downstream from the turbulence grid.

In the velocity measurements, a DANTEC laser-Doppler velocimetry (LDV) system consisting of a fiber-optic probe head and a 310mm focal-length lens were employed to obtain a streamwise velocity at the $760\delta$ downstream from the turbulence grid. The laser light was green beams with wavelengths of 514.5 nm, and the distance between two beams was 38 mm. Signal analysis was accomplished by BSA with FFT. The tracer particle seeded flow was acrylic colloid. The green beams passed through the $y$ direction. The measuring volumes were $0.113 \times 0.951$ mm$^2$. We carried out the LDV measurement of the streamwise velocity at the channel center for $1100 \leq Re_m \leq 2400$.

## 3. Result and Discussion
### 3.1 Flow visualization of the turbulent stripe structure

Figure 2 shows flow visualization with flakes. The dimension of the field shown here is $126\delta \times 84\delta$. As for the image, the mean value of 1800 images of the brightness of each pixel was subtracted. When the direction of a platelet is completely random, the light is hardly reflected. There is the turbulent region which can be marked by black and uniform region. In contrast, when the platelet aligns in one direction, strong reflection will be observed at the picture shown the white region. In a quasi-laminar region, the reflected light intensity was high, and longitudinal streaks which reflected light efficiently were observed. In a turbulent region, the light intensity was smaller than the quasi-laminar region. Although the picture is not shown here, a trial to illuminate the flakes by laser light revealed that the flakes were blinking in the dark region of figure 2, which corresponds to the turbulent fluctuation. The flow state at $Re_m = 2400$ was homogeneous turbulence (i.e., typical high-Reynolds-number turbulence), as shown in figure 2(a). This flow lacks any apparent large-scale structure, while the dominant visible length-scale was smaller than the channel width. Turbulent stripe occurred for $1700 \leq Re_m \leq 2000$. When $Re_m$ was decreased down to 2000, the turbulent motions induced by the turbulence grid are locally suppressed, resulting in a turbulent-laminar banded pattern, as shown in figure 2(b). We called this pattern a "turbulent stripe" structure. The stripe pattern was maintained at $Re_m = 1850$ and 1700, and its size was slightly larger than that observed at $Re_m = 2000$: see figures 2(c) and 2(d). However, the stripe shape at $Re_m = 1700$ was almost broken down, and seemed likely a number of turbulent spots arranged in a staggered pattern. Therefore it can be suggested that the turbulent stripe seems to be configured with combination of multiple turbulent spots. The stripe shape completely broke down for $Re_m < 1700$. One or two turbulent spots were observed to occur at $Re_m = 1600$, as shown in figure 2(e). As just described, a sequence of transitions between a sufficiently high-Reynolds-number flow and a low-Reynolds-number flow, in which turbulent spots occur spontaneously, is schematically illustrated in figure 3. At $Re_m = 1300$, the flow was stable and laminar. Consequently, the critical Reynolds number was $Re_m = 1300$.

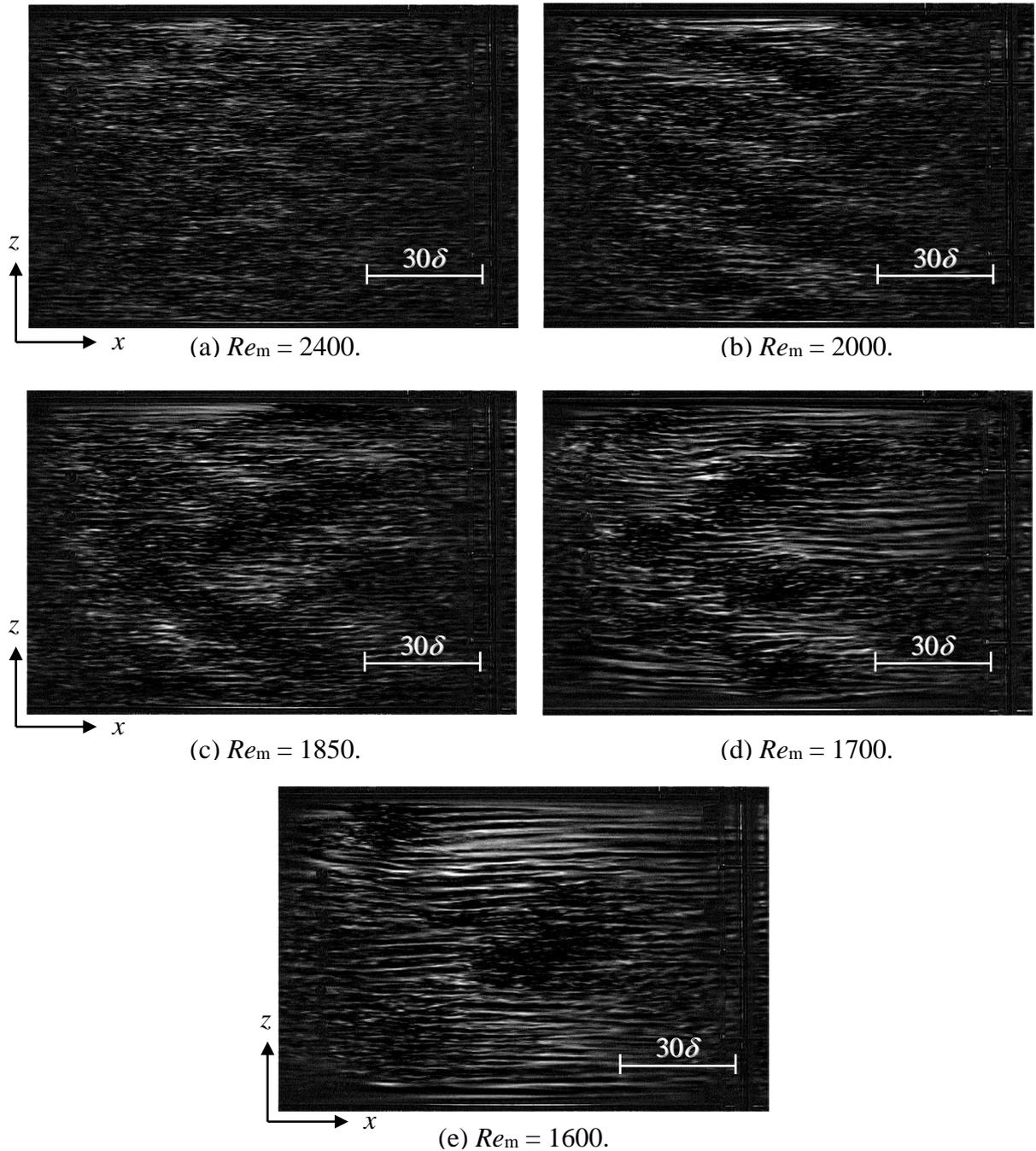

Figure 2: Flow visualization with flakes at each Reynolds number. The field shown here is $126\delta \times 84\delta$. Mean flow direction is from left to right.

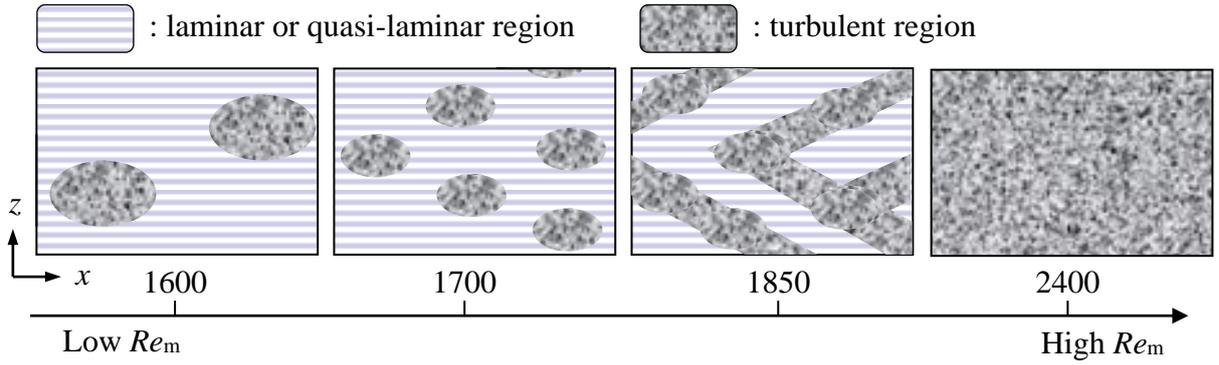

Figure 3: Schematic illustrated transitional channel flow fields from low Reynolds number to sufficiently high Reynolds number.

Small streaks, which were visible in turbulent regions at $Re_m = 2400$ as shown in figure 2(a), corresponded to intrinsic streaky structures in the high-Reynolds-number wall turbulence. On the other hand, for $Re_m = 2000 \sim 1600$, the visible streaks in quasi-laminar regions were different from those in the turbulent regions. The longitudinal range of the streaks in quasi-laminar regions was much longer than the streaks in turbulent regions. These streaks indicate longitudinal vortices and these flows in light regions are unstable flows. These streaks were only observed in transitional channel flow. Thus, the light regions that different from turbulent and laminar regions are quasi-laminar regions.

Figure 4 is a typical flow visualization image of the turbulent stripe at $Re_m = 2000$ in $(x, z)$ plane. The dimension of the field shown here is $70\delta \times 70\delta$. As for the image, the mean value of 300 images of the brightness of each pixel was subtracted. The inclination angle $\theta$ was an angle of a stripe structure with the streamwise direction. Each positive and negative angle of the stripe pattern had an equal chance to occur. The streamwise wavelengths $\lambda_x$ and the spanwise wavelengths $\lambda_z$ were a pair of turbulent region and laminar region in the streamwise direction and spanwise direction, respectively. In this work, these values of $\theta$, $\lambda_x$ and $\lambda_z$ were obtained from the flow visualization. The turbulent band was inclined about $|\theta| = 20º \sim 30º$ against the streamwise direction. The streamwise wavelength $\lambda_x$ was about $60\delta$. The spanwise wavelength $\lambda_z$ was about $20\delta \sim 30\delta$. The streamwise length of the quasi-laminar regions were about $30\delta$ and the spanwise lengths of the quasi-laminar regions were about $10\delta \sim 20\delta$.

Figures 5 and 6 present streamwise and spanwise pre-multiplied energy spectrum $k_i E(k_i)$ of light intensity, as a function of wavelength $\lambda_i$. In each figure, wavelengths on the abscissa in (a) and (b) are normalized by the viscous length (i.e., in wall units) or by the channel half

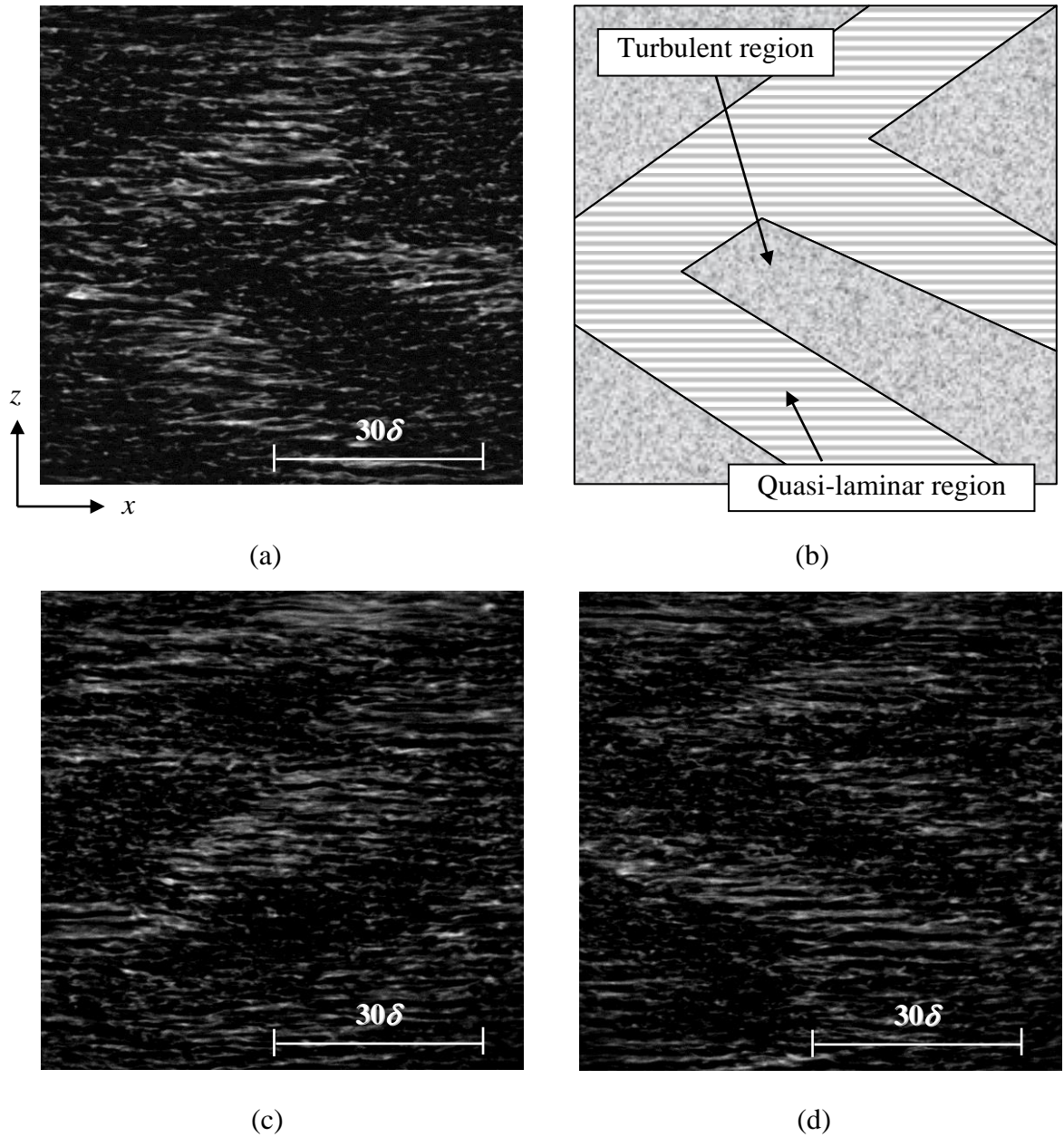

Figure 4: Flow visualization with flakes at $Re_m = 2000$. The field shown here is $70\delta \times 70\delta$. Mean flow direction is from left to right. In (a),(c), and (d), flow visualization with flakes. In (b), a schematic diagram of turbulent stripes extracted from the flow visualization displayed in (a).

width, respectively. Note that a peak of a pre-multiplied energy spectrum is known as a mean distance of spatial structures in a relevant direction [12]. In figure 5(a), the peak of pre-multiplied energy spectrum near $\lambda_x^+ = 350 \sim 500$ ($\lambda_x/\delta = 4 \sim 7$) corresponds the vortex in turbulent regions. Here, the superscript (+) denotes a quantity non-dimensionalized by the friction velocity $u_\tau$ and $\nu$, namely in wall units. As $Re_m$ decreased, the peak position shifts slightly to longer wavelengths. In figure 5(b), the peak at $\lambda_x/\delta = 35$ (around $\lambda_x^+ = 2300$) corresponds to the longitudinal range of the streaks in quasi-laminar regions. As $Re_m$ decreased, the most energetic wavelength shifts from the peak around $\lambda_x^+ = 350 \sim 500$ to $\lambda_x/\delta$

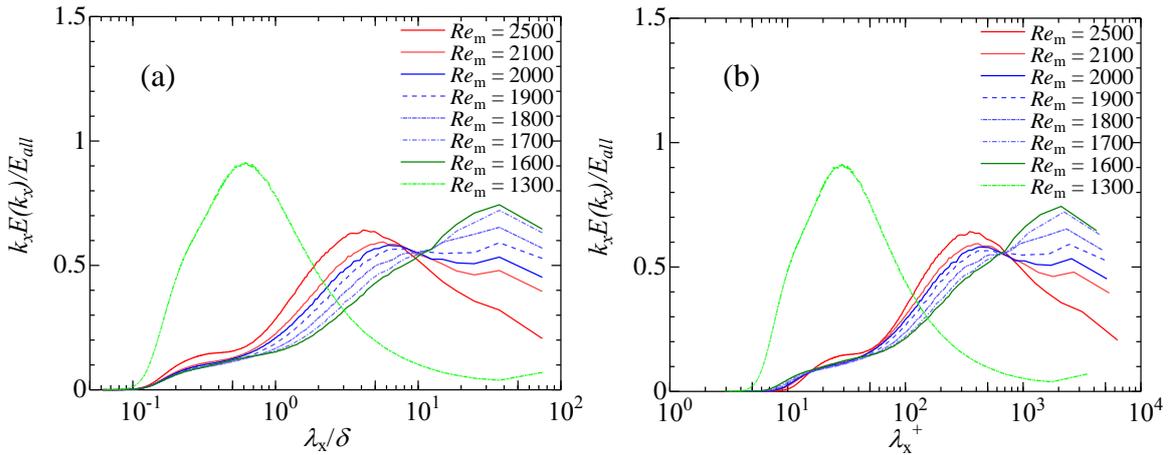

Figure 5: Pre-multiplied energy spectrum of light intensity as a function of streamwise wavelength scaled by (a) channel half width $\delta$, or (b) in wall unit.

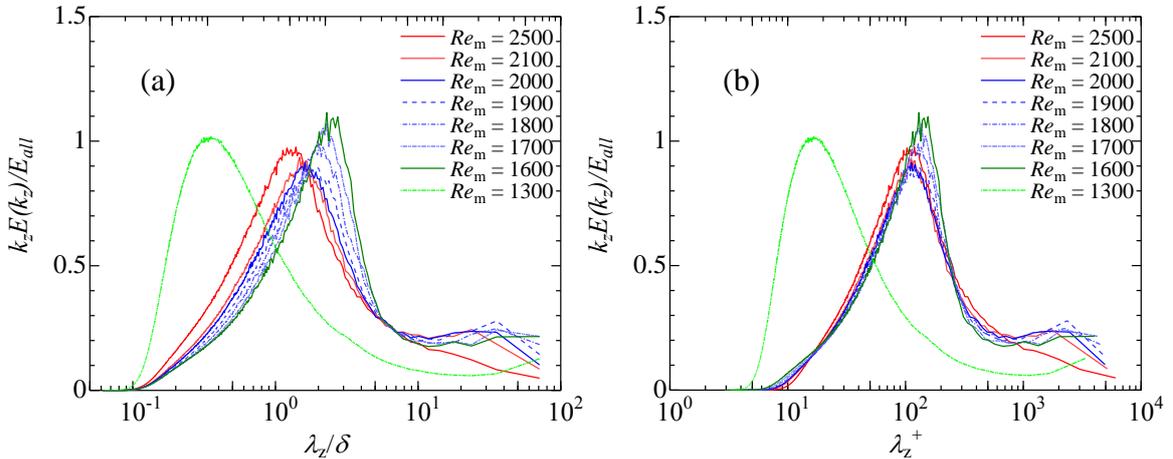

Figure 6: Same as figure 5 but for spanwise wavelength.

= 35. In figure 6(a), the peak near $\lambda_z^+ = 100$ ($\lambda_z/\delta = 1.6$) corresponds to distance of spanwise direction of streaks. As $Re_m$ decreased, the peak position shifts slightly to longer wavelengths. In other words, with the decrease of $Re_m$, the spanwise distances of streaks become large. However, the peak of spanwise direction near $\lambda_z^+ = 2300$ ($\lambda_z/\delta = 35$) was not clear. The peaks of streamwise direction at $\lambda_x/\delta = 35$ were observed for each $Re_m$ (see figure 5 (a)). Although the DNS result by Tsukahara *et al.* [5] indicated that longitudinal streaks in the quasi-laminar region were enlarged at lower $Re_m$, the streaks were observed to be rather constant in the present experiment. Thus, it is difficult to estimate quantitatively such long wavelength in this case.

### 3.2 Comparison of turbulent-stripe structure between experiment and DNS

Table 1 presents the $Re_m$ range, $|\theta|$, $\lambda_x$, and $\lambda_z$ for which turbulent stripes have been observed in the present experiment and the DNS performed by Tsukahara *et al.* [5]. Also shown are the streamwise and spanwise lengths of the quasi-laminar regions. The spanwise spacing of the streaky structure in the quasi-laminar regions was also compared. The size and angle of the stripe structure observed experimentally show a good agreement with the DNS results. However, a considerable discrepancy was found in the $Re_m$ range: the range detected by the DNS was wider than that by the present experiment. The upper bound of $Re_m$ for the experiment was lower than DNS since there exist disturbances, such as roughness in the channel surface, in the experiment. The slightly large value of the lower bound of $Re_m$ was caused by the sidewall of the channel, especially limited spanwise extent. The spanwise channel size of 80$\delta$ is not large enough for the turbulent stripe to occur at the lower Reynolds number than 1700. Similar tendency was already found in the DNS [3], revealing that turbulent motions were prone to decay even at $Re_m$=2300 if the computational domain was smaller than the intrinsic turbulent structures.

Table 1: Comparison of turbulent stripe structure between experiment and DNS: $\theta$, angle of the streamwise direction; $\lambda_x$, the streamwise spacing of a pair of laminar-turbulent band.

|  | Present Exp. | DNS [5] |
|---|---|---|
| $Re_m$ range of occurring turbulent stripe | 1700 ~ 2000 | 1570 ~ 2330 |
| $|\theta|$(deg) | 20° ~ 30° | 20° ~ 25° |
| $\lambda_x$ | 60$\delta$ | 66$\delta$ |
| $\lambda_z$ | 20$\delta$ ~ 30$\delta$ | 22$\delta$ |
| Streamwise length of the quasi-laminar region | 30$\delta$ | 30$\delta$ |
| Spanwise length of the quasi-laminar region | 10$\delta$ ~ 20$\delta$ | 10$\delta$ |
| Spanwise spacing of streaks in quasi-laminar in wall units | 100 $\nu/u_\tau$ | 100 $\nu/u_\tau$ |

### 3.3 Comparison of similar phenomena in other flow systems

The turbulent stripe in plane Couette flow was studied by Prigent *et al.* [9]. The turbulent stripes in plane Couette flow was observed in the transition from turbulent to laminar. However, it was not certain that the laminar regions of the turbulent stripe in plane Couette flow were different from the quasi-laminar regions. We extracted the angle of the turbulent stripe in plane Couette flow against the streamwise direction on their flow visualizations. The turbulent stripes in plane Couette flow was inclined about $|\theta| = 20°$~30°. The angle of the turbulent stripe in Poiseuille flow was good agreement with turbulent stripe in plane Couette flow. The turbulent stripe in plane Couette flow exhibited a spatial periodicity of the laminar region and turbulent region in the streamwise direction of the order of 50 times gap between the walls. The streamwise wavelengths $\lambda_x$ of turbulent stripe in Poiseuille flow was smaller than that in plane Couette flow.

The spiral turbulence was studied by Andereck *et al.* [7] and Prigent *et al.* [9]. Andereck *et al.* observed spiral turbulence in counter-rotating cylinders. Transitions between states were determined as functions of the inner- and outer-cylinder Reynolds numbers, $R_i$ and $R_o$. The laminar spirals were observed in low $R_i$ and high $R_o$ (e.g. $R_i = 240$ and $R_o = -300$). When increasing $R_i$, the intermittent turbulent spots were observed in a flow with laminar spirals (e.g. $R_i = 590$ and $R_o = -1500$). On the other hand, the turbulent spots in Poiseuille flow were observed in quasi-laminar regions (see figure 5(e)). Thus, the streaks in quasi-laminar regions were similar to the laminar spirals in Taylor-Couette flow. When increasing $R_i$, spiral turbulence, which consists of a periodic alternation of spiral turbulent and laminar spiral, was observed (e.g. $R_i = 1000$ and $R_o = -3500$). Both the turbulent stripe and spiral turbulence are composed of fully turbulent region and quasi-laminar region. And the quasi-laminar region is not completely laminar, but it contains longitudinal vortices. This feature is commonly observed in above two flows.

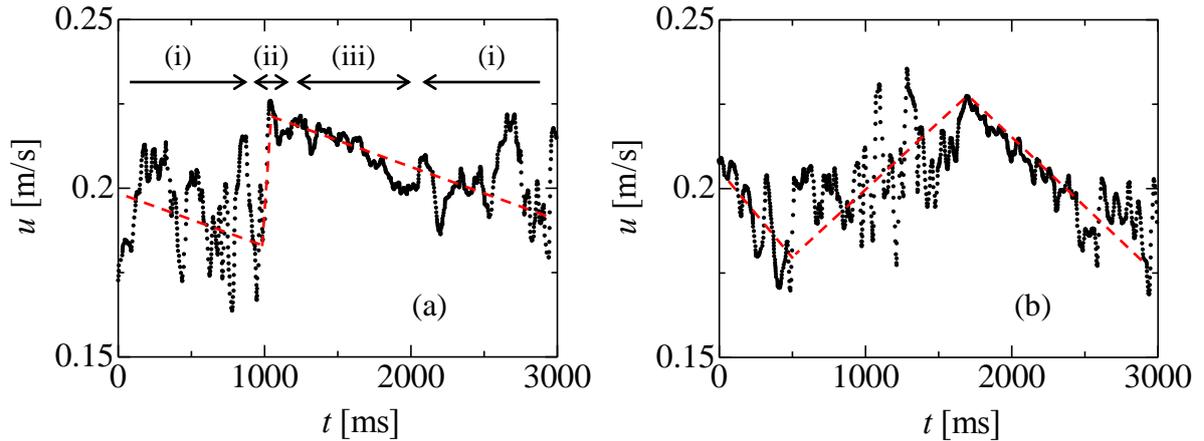

Figure 6: Time trace of streamwise velocity measured by LDV at $Re_m = 2000$. The measurement point is the channel center at $760\delta$ downstream from the entrance. (a) Pattern of puff-like time trace, and (b) another pattern.

In conclusion, the comparison of the turbulent stripe in the channel flow and other flow systems revealed that the turbulent stripes in the plan Poiseuille flow as well as the in plane Couette flow and the spiral turbulence in Taylor-Couette flow have a common feature. The Taylor-Couette flow has a strong effect of centrifugal force, so this force can affect the flow stability. It is interesting that the common phenomenon was observed in a plane Couette flows which has no effect of centrifugal force.

### 3.5 Time trace of the streamwise velocity

Figures 6 shows representative time trace of the streamwise velocity measure by LDV at the channel center for $Re_m = 2000$. These time traces are similar to those in the puff in a pipe [1]. As can be seen in figure 6(a), the highly disordered turbulent region has appeared intermittently (region-(i) in the figure). Then, the velocity increased rapidly (ii), and the velocity decreased gradually with small fluctuations (iii). After that, the velocity fluctuation became large again (region-(i)). Hence, the boundary from the turbulent region to the quasi-laminar region was steep, and that from turbulent region to the quasi-laminar region was not clear. This pattern was occasionally observed for $1700 \leq Re_m \leq 2000$. In addition, figure 6(b) shows a slightly different time trace at $Re_m = 2000$. The velocity increased gradually with large fluctuations and decreased gradually with small fluctuations. Both boundaries from turbulent to quasi-laminar regions and from quasi-laminar to turbulent regions were moderate. This pattern was also sometimes observed. The turbulent stripe seemed to be configured with combination of a number of turbulent spots (see figure 3(d)). Thus, the turbulent stripe seemed to have the particularly low speed regions and not. The pattern of puff-like time trace was measured in the particularly low speed regions. On the other hand, another pattern was measured in the other regions. This is required for further study.
## 4. Conclusion
We performed the flow visualization and the LDV measurement to examine the turbulent stripe in a channel flow at transitional Reynolds number of $Re_m$ = 1400 ~ 2400. Based on the present experiment, we can draw the following features with respect to the turbulent stripe:
- angle against the streamwise direction was $\theta$ = 20º ~ 30º,
- spatial period of the stripe pattern in the streamwise direction was $\lambda_x = 60\delta$,
- spatial period of the stripe pattern in the spanwise direction was $\lambda_x = 20 ~ 30\delta$,
- the spanwise spacing of streaks in quasi-laminar regions was about $1.6\delta$,
- time trace of steamwise velocity was similar to the puff in a transitional pipe flow.

In addition, the critical Reynolds number was found to be $Re_m$ = 1300.

This paper is a revised and expanded version of a paper entitled "An Experimental Study on Turbulent-Stripe Structure in Transitional Channel Flow", presented by S. Hashimoto, A. Hasobe, T. Tsukahara, Y. Kawaguchi, and H. Kawamura, at the Sixth International Symposium on Turbulence, Heat and Mass Transfer, Rome, Italy, 14–18 September 2009.